\documentclass{aa}
\usepackage{natbib}
\usepackage[dvips, pdftex]{graphicx}
\usepackage{epstopdf}
\usepackage{txfonts}
\bibpunct{(}{)}{;}{a}{}{,}

\begin{document}
\authorrunning{Hu et al.}

   \title{Gravitational settling in pulsating subdwarf B stars \\and their progenitors}

   \author{Haili Hu
          \inst{1,2}
          \and
          E.~Glebbeek\inst{3}                
          \and
          A.~A.~Thoul\inst{4}  
           \and               
          M.-A.~Dupret\inst{4}     
          \and                                 
          R.~J.~Stancliffe\inst{5}
          \and
          G.~Nelemans\inst{1}
          \and             
          C.~Aerts\inst{1,2}                                                                                     
         }

   \offprints{hailihu@astro.ru.nl}

   \institute{Department of Astrophysics, IMAPP, Radboud University Nijmegen, PO Box 9010, 6500 GL, Nijmegen, the Netherlands
         \and
             Institute of Astronomy, Katholieke Universiteit Leuven, Celestijnenlaan 200D, 3001 Leuven, Belgium
         \and
        Department of Physics and Astronomy, McMaster University, 1280 Main Street West, Hamilton, Ontario, Canada L8S 4M1
         \and                
         Institute d'Astrophysique et G\'eophysique, universit\'e de Li\`ege, Belgium         
         \and                            
         Centre for Stellar and Planetary Astrophysics, Monash University, PO Box 28M, Clayton VIC 3800, Australia
             }

   \date{Received 7 April 2009 / Accepted 15 December 2009 }

 
  \abstract
   {Diffusion of atoms can be important during quiescent phases of stellar evolution. Particularly in the very thin inert envelopes of subdwarf B stars, diffusive movements will considerably change the {envelope structure and the} surface abundances on a short timescale. Also, the subdwarfs will inherit the effects of diffusion in their direct progenitors, namely giants near the tip of the red giant branch. This will influence the global evolution and the pulsational properties of subdwarf B stars.}
   {We investigate the impact of gravitational settling, thermal diffusion and concentration diffusion on the evolution and pulsations of subdwarf B stars. Although radiative levitation is not explicitly calculated, we evaluate its effect by approximating the resulting iron accumulation in the driving region. This allows us to study the excitation of the pulsation modes, albeit in a parametric fashion. Our diffusive stellar models are compared with models evolved without diffusion. 
   }
   {We use a detailed stellar evolution code to solve simultaneously the equations of stellar structure and evolution, including the composition changes due to diffusion. The diffusion calculations are performed for a multicomponent fluid using diffusion coefficients derived from a screened Coulomb potential. We constructed subdwarf B models with a mass of 0.465 M$_{\odot}$ from a 1 M$_{\odot}$ and 3 M$_{\odot}$ zero-age main sequence progenitor. The low mass star ignited helium in {an energetic} flash, while the intermediate mass star started helium fusion gently. For each progenitor type we computed series with and without atomic diffusion.
   }
   {Atomic diffusion in red giants causes the helium core mass at the onset of helium ignition to be larger. We find an increase of 0.0015 M$_{\odot}$ for the 1 M$_{\odot}$ {model} and 0.0036 M$_{\odot}$ for the 3 M$_{\odot}$ {model}.  The effects on the red giant surface abundances are small after the first dredge up. The evolutionary tracks of the diffusive subdwarf B models are shifted to lower surface gravities and effective temperatures due to outward diffusion of hydrogen. This affects both the frequencies of the excited modes and the overall frequency spectrum. Especially the structure and pulsations of the post-non-degenerate sdB star are drastically altered, proving that atomic diffusion cannot be ignored in these stars.  Sinking of metals could to some extent increase the  gravities and temperatures due to the associated decrease in the stellar opacity. However, this effect should be limited as it is counteracted by radiative levitation. }
{}


   \maketitle

\section{Introduction}
Atomic diffusion in stars can change the chemical abundances during different evolutionary stages. For example, the chemically peculiar subdwarf B (sdB) stars show surface abundance anomalies most likely caused by diffusive processes \citep{fontaine1997}. These stars, which are believed to be low mass ($\sim$0.5 M$_{\odot}$) core-He-burning stars, have atmospheres typically dominated by H and deficient in He \citep{heber1991}. In addition, \citet{michaud2007} found significant differences between stellar models evolved with and without diffusion on the Red Giant Branch (RGB) and the Horizontal Branch (HB). Some of their results should also apply to sdB stars, which are commonly believed to be Extreme Horizontal Branch (EHB) stars and the descendants of red giants.

Some sdB stars show pulsations, and two types of pulsators are distinguished; the short-period variable EC 14026
stars \citep{kilkenny1997}, and the long-period variable PG 1716 stars
\citep{green2003}. The rapid oscillations in EC 14026 stars are interpreted as low-order $p$-modes, driven by the
$\kappa$-mechanism operating in the iron opacity bump \citep{charpinet1996}. The same mechanism has been shown to excite long-period, high-order $g$-modes in models slightly cooler than the PG 1716 stars \citep{fontaine2003}. This temperature discrepancy might be related to the stellar opacities \citep{jeffery2006b}. The local iron enhancement necessary in the driving region around $\log T\approx 5.3$ is caused by the competing diffusion processes of radiative levitation and gravitational settling. 

Despite the obvious importance of diffusion in these stars, sdB evolutionary models did not include this process consistently up to now. Instead, it was often assumed that, as a result of gravitational settling, the thin envelope is H-rich and the H-profile is arbitrarily steep. However, the envelope composition and the precise form of the He-H transition layer are important for the pulsation modes. 
Furthermore, the assumption of such a H-envelope is inappropriate for the newly proposed post-non-degenerate sdB stars {by \citet{han2002}}. \citet{hu2008} showed that {these stars} have a broad He-H transition layer extending to {deeper regions}, where diffusion is not expected to work efficiently. This is in contrast to the canonical post-He-flash sdB models, that  inherit a steep H-profile from their progenitors on the RGB. 

While the inclusion of gravitational settling, thermal diffusion and concentration diffusion during stellar evolution is relatively straightforward as we show here, the implementation of radiative levitation is far more difficult. Still, radiative levitation has been treated consistently in some stellar evolutionary models, for example for Population I and II main sequence stars \citep{richard2001, richard2002} and HB stars \citep{michaud2007}. These studies have led to important insights in the stellar structure, evolution and abundance anomalies, see \citet{michaud2008} for a review. Although it is eventually also our aim to include radiative accelerations in our sdB models, it is beyond the scope of this work.

In order to study the mode excitation in sdB stars nevertheless, different approximations have been made in the literature. One could simply enhance Fe uniformly through the entire stellar envelope \citep{charpinet1996, jeffery2006a}, but then the stellar evolution and pulsations would be unrealistically altered. A more sophisticated treatment assumes a diffusive equilibrium profile of Fe, see \citet{charpinet1997}. These authors argued that the equilibrium state is reached on timescales much shorter than the sdB evolutionary  timescale {($\sim$10$^5$ {compared to} $\sim$10$^8$ years)}. \citet{hu2008} approximated the equilibrium Fe profile with a Gaussian function centered around $\log T=5.3$. Such a parametric approach is convenient for the study of mode excitation, because the exact shape of the non-uniform Fe profile does not greatly affect the driving, as long as the resulting  opacity bump is large enough. Secondly, Fe is only enhanced in the driving region, corresponding to a very thin mass-shell ($\lesssim$$10^{-6}$ M$_{\odot}$). Thus, such a parametric approach does not model the shape of the Fe distribution in detail, but is more realistic than simply increasing the metallicity in the entire envelope.

In this study, we constructed sdB models with gravitational settling, temperature diffusion and concentration diffusion from the zero-age main sequence (ZAMS) to the EHB. 
Radiative accelerations were neglected, but in order to evaluate how our results might be affected, we  enforced Fe enhancement in one series of calculations, in the same way as in \citet{hu2008}. 
For convenience, we will refer to the collective effects of gravitational settling, thermal and concentration diffusion as `atomic diffusion' in the rest of this paper. The work presented here should be considered as a necessary but intermediate step towards achieving our goal of improving the seismic models for sdB stars. A more detailed study, including the effects on sdB pulsations, will be presented in a future paper as we are currently implementing additional transport processes in our models (e.g., radiative levitation, thermohaline mixing, and turbulence). Here, we focus on the implementation of atomic diffusion and the qualitative effects on the evolution tracks and stability of the pulsation modes of sdB stars.

In the next section, we introduce the method followed in this work.  In Section \ref{4diffusion} we present the diffusion equations. In Section \ref{4evolution}, we describe the stellar evolution code and the input physics. In Section \ref{4sun}, we show calibrated solar models as  a test for the updated diffusion and evolution code. In Section \ref{4rgb}, we examine the effects of atomic diffusion in red giants, which we used to build diffusive sdB models presented in Section \ref{4sdb}. The conclusions are discussed in Section \ref{4concl}.
  
\section{Method}\label{4method}  

\subsection{The diffusion equations}\label{4diffusion}  
The effects of diffusion are in general small, and this has led to various, sometimes unnecessary, approximations in its treatment, e.g.~considering trace elements and binary mixtures. Also, the calculation of the diffusion coefficients is greatly simplified by describing the interaction between particles by a pure Coulomb potential with a long range cut-off. However, a screened Coulomb potential gives more accurate results, especially at high densities \citep{paquette1986}. 
Here, we calculate the diffusion velocities due to gradients of pressure (i.e.~gravitational settling), temperature, and concentration in a multicomponent fluid. We use the routine by \citet{thoul1994} (hereafter TBL) to solve Burgers flow equations \citep{burgers1969}. The original TBL routine uses diffusion coefficients derived from a pure Coulomb potential with a cut-off at the Debye length. We have updated it to make use of more accurate diffusion coefficients derived from a screened Coulomb potential by \citet{paquette1986}. 

For a review of the basic equations describing atomic diffusion in stars we refer to \citet{thoul2007}. We briefly recall the most important ones. The Burgers diffusion equations are ($N$ equations),
\begin{equation}\label{4difeq}
\frac{\textrm{d}p_i}{\textrm{d}r}+\rho_ig-n_iZ_ieE=\sum_{j\neq i}^N K_{ij}(w_j-w_i)+\sum_{j\neq i}^{N}K_{ij}z_{ij}\frac{m_jr_i-m_ir_j}{m_i+m_j},
\end{equation}
including the heat flow equation ($N$ equations),
\begin{eqnarray}\label{4heat}
\frac{5}{2}n_i k_{\rm B}\nabla T &=&\frac{5}{2}\sum_{j\neq i}^Nz_{ij}\frac{m_j}{m_i+m_j}(w_j-w_i)-\frac{2}{5}K_{ii}z_{ii}''r_i \nonumber\\
&&-\sum_{j\neq i}^N\frac{K_{ij}}{(m_i+m_j)^2}(3m_i^2+m_j^2z_{ij}'+0.8m_im_jz_{ij}'')r_i\nonumber\\
&&+\sum_{j\neq i}^N\frac{K_{ij}m_im_j}{(m_i+m_j)^2}(3+z_{ij}'-0.8z_{ij}'')r_j.
\end{eqnarray}
In addition, we have two constraints: current neutrality,
\begin{equation}\label{4current}
\sum_i Z_i n_i w_i = 0,
\end{equation}
and local mass conservation,
\begin{equation}\label{4mass}
\sum_i m_i n_i w_i = 0.
\end{equation}
In the above $2N+2$ equations, $p_i$, $\rho_i$, $n_i$, $Z_i$, and $m_i$ denote the partial pressure, mass density, number density, charge and mass for species $i$, respectively. $N$ is the total number of species including electrons. The $2N+2$ unknown variables are the $N$ diffusion velocities $w_i$, the $N$ heat fluxes $r_i$, the gravitational acceleration $g$, and the electric field $E$. We take the diffusion coefficients, $K_{ij}$, $z_{ij}$, $z_{ij}'$ and $z_{ij}''$, from \citet{paquette1986}. The system of Eqs.~(\ref{4difeq}) - (\ref{4mass}) can be written as a matrix equation, and it can then be solved by LU decomposition, i.e. by decomposing the matrix in a lower and upper triangular matrix. The procedure we followed is described in detail by TBL. We note that TBL eliminated the concentration gradient of He by demanding charge neutrality. However, this is unnecessary and we find it more convenient to keep all ionic concentration gradients. 

A rigorous study requires the calculation of the degree of ionization, and the treatment of each ion as a separate species. As  a simplification, one often uses the mean ionic charge. However, we found that this gives inconsistent results in regions where the degree of ionization changes rapidly.  Also, difficulties arise in the case of neutral atoms, for TBL defined the concentrations with respect to electrons. Furthermore, we have not included diffusion coefficients of neutral atoms, which are due to atomic polarizability rather than Coulomb scattering. To avoid these complications, we simply assume full ionization for the calculation of the diffusion velocities, although we are aware that this might lead to a slight underestimation of the diffusion velocities \citep{turcotte1998, schlattl2002}. For the equation of state the ionization of H, He, C, N, O and Ne are calculated using an approximate pressure ionization model \citep{pols1995}.

The diffusion velocity can be expressed in terms of gradients of ion abundances, pressure and temperature, 
\[
w_i = \sum_j a_{ij} \frac{\partial \ln  X_j}{\partial r}+a_P\frac{\partial \ln P}{\partial r}+a_T\frac{\partial \ln T}{\partial r},
\]
where the coefficients $a_{ij}$, $a_P$ and $a_T$ follow from the TBL procedure.
Having calculated the diffusion velocities, we can now solve the equations for the composition changes,
\begin{equation}\label{4dxdt}
\frac{\partial X_i}{\partial t} + R_{\rm nuc, i}= \frac{1}{\rho r^2}\frac{\partial}{\partial r}\Big(
\sigma\rho r^2\frac{\partial X_i}{\partial r}\Big) - \frac{1}{\rho r^2}\frac{\partial}{\partial r}(\rho r^2 X_i w_i),
\end{equation}
where $R_{\rm nuc, i}$ is the rate of change of species $i$ due to nuclear reactions, $\sigma$ is the turbulent diffusion coefficient for convective mixing, and the last term gives the composition changes due to atomic diffusion. 

\subsection{The stellar evolution code and input physics} \label{4evolution} 
The stellar models are constructed with the stellar evolution code STARS developed by
\citet{eggleton1971, eggleton1972, eggleton1973}, and \citet{ faulkner1973}. This code has been updated by \citet{pols1995} to make use of an equation of state that includes pressure ionization and Coulomb interaction. \citet{stancliffe2008} implemented gravitational settling in STARS by considering trace elements in a H background. Here, we solve the full set of Burgers equations as described in the previous section.

\citet{stancliffe2006} distinguishes three different methods for a stellar evolution code to solve the structure equations together with the equations for composition changes: the fully simultaneous, partially simultaneous and non-simultaneous approach. The STARS code follows the fully simultaneous approach, i.e. the composition equations are solved simultaneously with the structure (and the mesh-spacing). While the non-simultaneous and the partially simultaneous approach can encounter numerical instabilities if the diffusion timescale becomes short, the simultaneous approach is not hampered by this. The drawback, however, is that it is computationally expensive to follow a large number of nuclear species in this way. We therefore limit to the abundance changes of seven species:  H, He, C, N, O, Ne, and a fictitious species gathering all the other ones. Thus, the heavy elements that are not explicitly followed are assumed to diffuse equally.

The code uses nuclear reaction rates from the NACRE compilation \citep{angulo1999} supplemented by reaction rates from \citet{caughlan1988}. The $^{14}$N(p,$\gamma$)$^{15}$O reaction rate is reduced to 0.64 times the NACRE value, as suggested by \citet{herwig2006} and \citet{formicola2002}. The neutrino loss rates are from \citet{itoh1989, itoh1992}. The metal mixture is scaled to solar abundances \citep{grevesse1993}. Convection is treated with the mixing-length prescription of \citet{bohm1958}. We use a mixing-length parameter (the ratio of the mixing-length to the local pressure scale height) of $\alpha=l/H_p=2.1$, which gives an excellent fit to the Sun with the input physics used here (see Section \ref{4sun}). 

Both convective and semi-convective mixing are treated as diffusion processes (see Eq.~\ref{4dxdt}). It is assumed that mixing occurs in regions where
\[
\nabla_{\rm rad}>\nabla_{\rm ad}-\delta_{\rm ov}/(2.5+20\beta+16\beta^2),
\]
where $\beta$ is the ratio of radiation pressure to gas pressure and
$\delta_{\rm{ov}}$ is the overshooting parameter. Convective overshooting is included  for stars with $M_{\rm ZAMS}> 1.5$ M$_{\odot}$, using an overshooting parameter $\delta_{\rm{ov}}=0.12$, which corresponds to an overshooting length of $\sim$$0.25 H_p$  \citep{schroeder1997, pols1997, pols1998}. For He-burning cores, convective overshooting is also included. The value of $\delta_{\rm{ov}}=0.12$ encompasses any additional mixing beyond the formal Schwarzschild boundary, regardless of whether it is caused by the actual overshooting of material from the convective region or whether it is caused by other processes, e.g.~rotationally induced mixing, atomic diffusion.
So when these mixing processes are explicitly modelled in the code, as atomic diffusion is now, the parameter $\delta_{\rm{ov}}$ should actually be redetermined.  Such a study is beyond the scope of this paper, and will be performed elsewhere (Glebbeek et al., in preparation).

Mass-loss on the RGB is described by a version of Reimers' law \citep{reimers1975} based on a physical approach by \citet{schroeder2005}:
\[
\dot{M}_{\rm{wind}}=\eta \frac{(R/R_{\odot})(L/L_{\odot})}{(M/M_{\odot})}
\Big(\frac{T_{\rm eff}}{4000\textrm{ K}}\Big)^{3.5}\Big(1+\frac{g_{\odot}}{4300g_*}\Big),
\]
with the parameter  $\eta$ set to $8\times10^{-14} \textrm{  M}_{\odot}\textrm{yr}^{-1} $. On the EHB, mass-loss is ignored in order to evaluate solely the effects of diffusion. It is expected that the rates in that phase are below $10^{-12}$ M$_{\odot}$yr$^{-1}$  \citep{unglaub2001, vink2002}. 

Due to diffusion, the H-abundance in the outer stellar layers will exceed the initial value. If the envelope is radiative and mass-loss is limited, the outer layer will consist of almost pure H. The  standard opacity tables in the STARS code \citep{pols1995, eldridge2004,chen2007} are not suitable here, because  they only go up to $X=0.8$. Furthermore, in the standard version of the evolution code, only the opacity table corresponding to the zero-age metallicity is loaded during an evolution run. 

Therefore, we constructed new opacity tables that are valid for H- and He-burning regions and for compositions up to $X=1$. We took the opportunity to implement the most recent values: OPAL opacities by \cite{iglesias1996} smoothly merged with the low-temperature opacities by \citet{ferguson2005}. These are then combined with updated conductive opacities \citep{cassisi2007} by reciprocal addition. The covered range is $2.7\leq\log T\leq 8.7$ and $-8 \leq \log R=\log \rho/T_6^3\leq 1$. At the high end of $\log T$ and $\log R$, where values are missing due to the non-rectangular form of the OPAL tables, we extrapolate linearly.

For interpolation convenience during the evolution calculations, the tables are made rectangular in ($\log T$, $\log \rho/T_6^3$,  $Z$, $XF$), where we defined the composition variable
\begin{eqnarray}\nonumber
XF = X - X_C-X_O. 
\end{eqnarray}
X is the hydrogen mass fraction,  $X_C$ and $X_O$ are the mass fractions of the enhanced carbon and oxygen, above that included in the metallicity. 
Thus, $XF$ is simply $X$ during H-burning, but follows CO enhancement during He-burning. We note that the low-temperature opacities by \citet{ferguson2005} do not include enhanced CO mixtures. Thus, our tables with CO enhancement are valid down to $\log T=4$, which is fine for sdB stars. 

We obtained tables for Z = $0$, $0.0001$, $0.0003$, $0.001$, $0.002$, $0.004$, $0.01$, $0.02$, $0.03$, and $0.04$. For each metallicity, we built tables for the compositions:  ($X$, $X_{C}+X_{O}$) = ($1-Z$, $0$), ($0.95$, 0), ($0.9$, $0$), (0.8, 0), (0.7, 0), (0.5, 0), (0.35, 0), (0.2, 0), (0.1, 0), (0, 0),  (0, $0.1+0$), (0, $0.3+0.1$), (0, $0.4+0.2$), (0, $0.4+0.4$), (0, $0.1+0.9-Z $). We derived the C/O ratio from a $0.47$ M$_{\odot}$ sdB star during He-burning. Due to the sensitive temperature dependence of the $3\alpha$ reactions, the C/O ratio is higher for more massive stars, but luckily, it is more or less constant within the narrow mass range of sdBs. Thus, we do not explicitly follow $X_C$  and $X_O$ separately (as in \citealt{eldridge2004}), but we use the fact that the C/O ratio is a function of  $X_{C}+X_{O}$ (as in \citealt{pols1995} and \citealt{chen2007}).

We also take into account the effect of metal diffusion on the opacity, and therefore interpolate in metallicity during the evolution. It should be noted that the opacity is still calculated for a fixed metal mixture \citep{grevesse1993}, except in the case that Fe is artificially enhanced in the sdB models. This is justified, since the heavy elements diffuse with roughly the same velocities.

The new opacity tables are written in a format suitable for the opacity routine that was originally developed for the CL\'ES evolution code by \citet{scuflaire2008}. This routine also calculates accurate opacity derivatives that are necessary for the stability analysis.

\section{Solar models}\label{4sun}
\subsection{Calibration of solar models}
An important test for a stellar evolution code is provided by the Sun. In the standard approach of calibrating solar models, the mixing length parameter $\alpha$, the initial He  and metal abundance are adjusted to produce, at the solar age ($4.57\times10^9$ yr, \citealt{bahcall1995}), models with observed solar radius ($6.9599\times10^{10}$ cm, \citealt{allen1973}), luminosity ($3.842\times10^{33}$ erg/s, \citealt{bahcall2001}) and surface metal fraction $Z_{\rm s}/X_{\rm s}$ (0.0245, \citealt{grevesse1993}). Because the luminosity of a stellar model is sensitive to the mean molecular weight, the He-abundance can be adjusted to yield the solar luminosity. The mixing length parameter determines the efficiency of energy transport by convection; at a fixed luminosity, a smaller $\alpha$ leads to a larger radius and thus a lower effective temperature. 

\citet{pols1995} found an approximate solar model, cooler and fainter within 0.2\% and 0.7\%, for  $\alpha=2.0$, $Y_{\rm i}=0.2812$, $Z_{\rm i}=0.0188$ with older opacity tables than used here \citep{rogers1992}, and no atomic diffusion. Since the stellar radius is also influenced by the opacity, stellar models using different opacity tables could require different values of $\alpha$ \citep{chieffi1995}. Furthermore, other physical inputs, such as the inclusion of atomic diffusion, will influence the calibration of solar models. Therefore, we perform a new solar calibration with the updated input physics described in Section \ref{4evolution}. 

We found for $\alpha=2.1$, $Z_i = 0.01928$, $Y _i = 0.2725$ a solar model with a radius and luminosity {both} within $0.01\%$ of the Sun. The surface mass fractions at the solar age are $Y_s = 0.248$ and $Z_{\rm s}/X_{\rm s}=0.0245$, which is consistent with \citet{grevesse1993}. The latest determination of $Z_{\rm s}/X_{\rm s}=0.0165$ by \citet{asplund2005}, although likely to be more accurate, poses a serious problem for helioseismology as pointed out by various authors, e.g.~\citet{serenelli2004}, \citet{montalban2004} and \citet{christensen-dalsgaard2009}. Also, the opacity tables use the \citet{grevesse1993} metal mixture. We have, therefore, not considered the new solar abundances here. 

The same ZAMS model evolved without atomic diffusion has at the solar age a radius and luminosity 1.6\% smaller than the diffusive model of the Sun. This can be understood in terms of the greater mean molecular weight in the core due to He-settling. Consequently, the  nuclear burning rate is higher, giving a larger radius and luminosity {for the diffusive solar model}. 

For the remainder of this paper, we use the mixing length parameter $\alpha=2.1$ for all our models. However, one should keep in mind that $\alpha$ could depend on the specific physical conditions, and could therefore vary with stellar mass and evolutionary phase.

\subsection{Diffusion velocities in the solar interior}
   \begin{figure}[b!]
   \begin{center}
  \includegraphics[angle=-90, width=9cm]{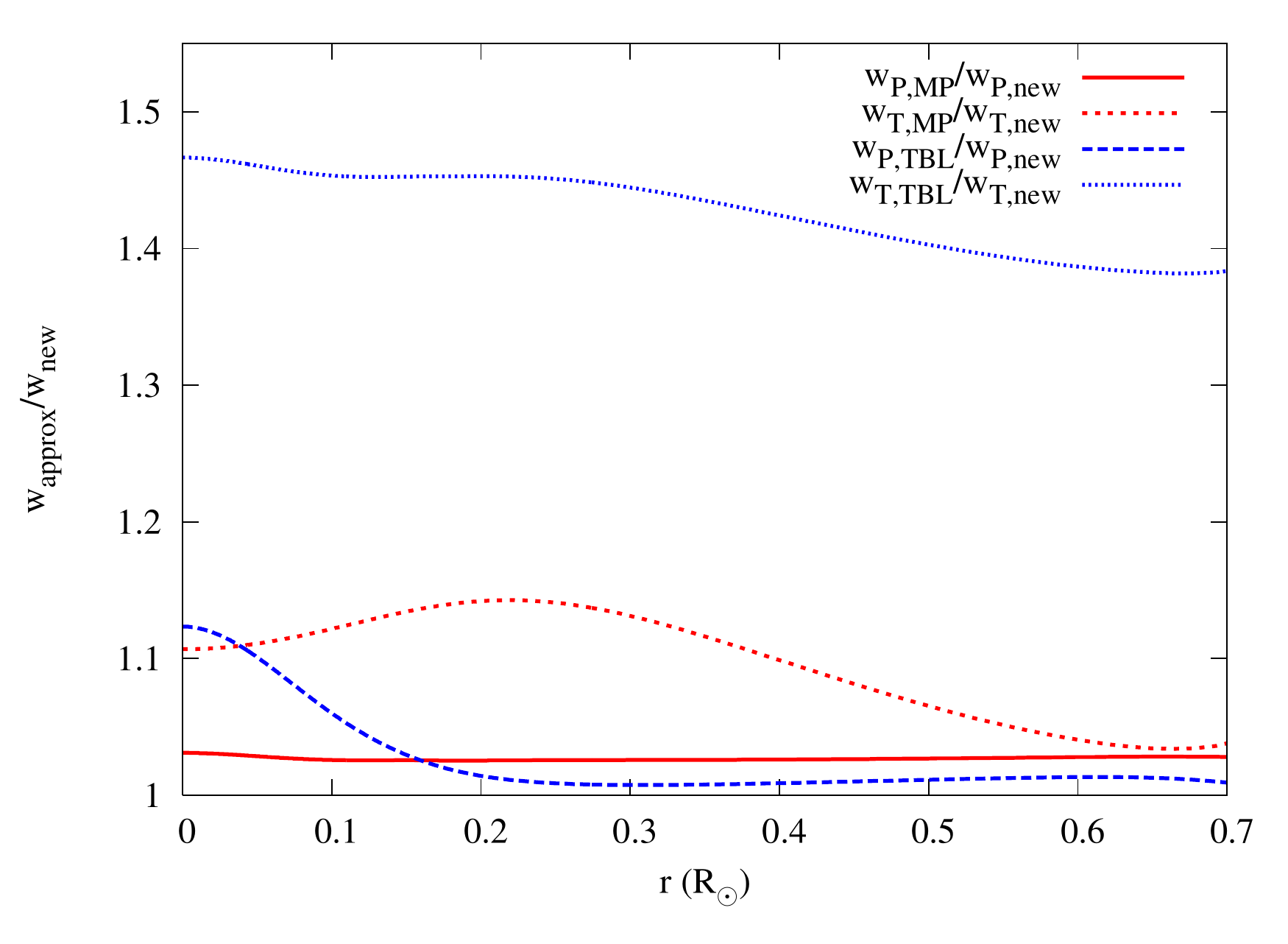}
   \caption{The ratio of the approximate diffusion velocities of H to our new exact values, as a function of the solar radius. The solid (MP) and long-dashed (TBL) lines give the ratio of the pressure terms, the short-dashed (MP) and dotted (TBL) lines indicate the temperature term ratios.}
              \label{4new_approx}
              \end{center}
    \end{figure}

It is illustrative to compare the diffusion velocities obtained with the updated diffusion routine to previous results, using (i) the simplified formulas by \citet{michaud1993} (hereafter MP) for diffusion velocities in a fully ionized  H-He mixture, and (ii) the original TBL routine. We calculated the H-diffusion velocity throughout the solar interior using Eq.~(17) of MP, where we used the standard solar model as described above. Fig.~\ref{4new_approx} shows the ratio of the MP H-diffusion velocities and our exact values. We evaluate the contributions due to pressure and temperature separately.  The ratio of the concentration gradients terms is not shown explicitly, because this almost coincides with the ratio of the pressure terms. Our results are also compared to values obtained with the original TBL routine. It is clear that the largest error in both the MP and the TBL approach is caused by the thermal diffusion term. It has already been noted by TBL that this error is likely caused by their assumption of $z_{ij}=0.6$. MP solved the Burgers equations for a H-He mixture analytically without the heat fluxes, and numerically with the heat fluxes. They then represented the effects of the heat fluxes by an ad hoc correction to the results obtained when neglecting those heat fluxes. It is, therefore, not surprising that their largest error is also in the thermal diffusion term. 

   \begin{figure*}[t!]
   \begin{center}
  \includegraphics[angle=-90, width=18cm]{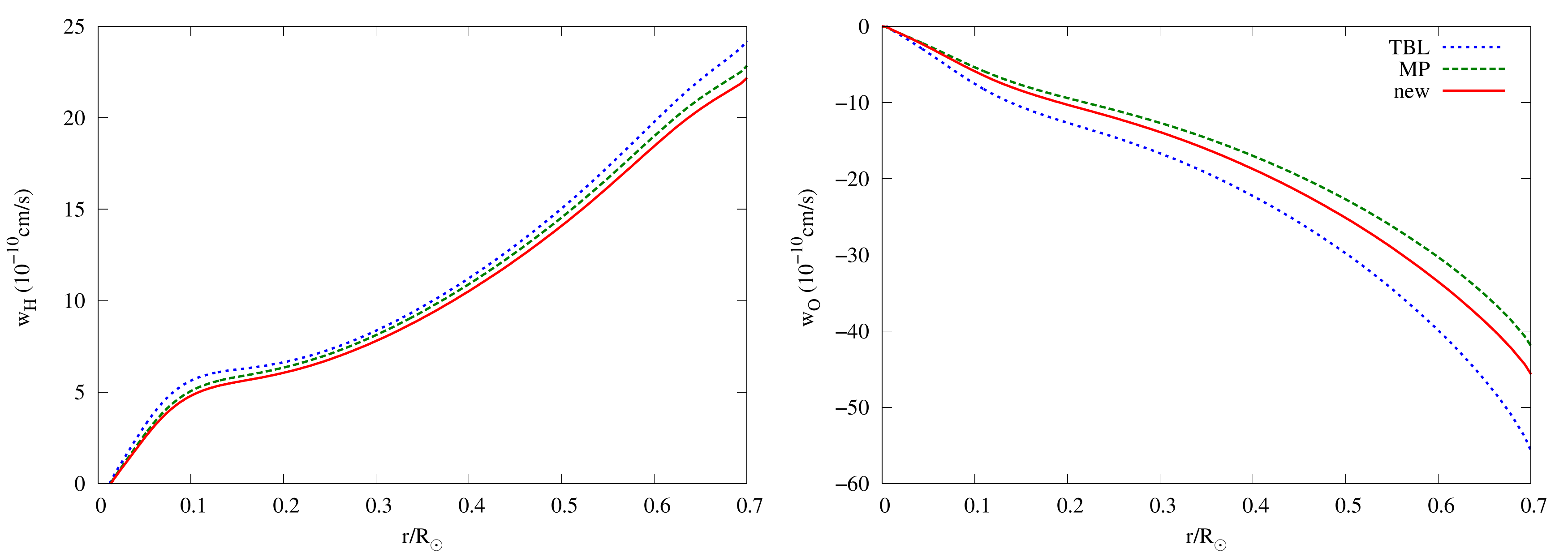}
   \caption{Diffusion velocities throughout the solar radiative interior. \emph{Left}: The total H diffusion velocity, and \emph{right}: the total oxygen diffusion velocity, using the three different approaches: TBL (dotted line), MP (dashed line) and our updated TBL approach (solid line).}
              \label{4wtot}
              \end{center}
    \end{figure*}

In  the left panel of Fig.~\ref{4wtot}, our total H-diffusion velocity is compared with results by MP and TBL. Our calculations give a slightly lower H-diffusion velocity. This is due to the overestimation of the thermal diffusion velocity in the approximations. A solar model evolved with the same input parameters as before, i.e.~$\alpha=2.1$, $Z_i = 0.01928$, $Y _i = 0.2725$, but using diffusion velocities from the original TBL routine, results in a radius larger by 0.09\% and is more luminous by 0.15\%. Although this effect is small, it becomes relevant within the desired accuracy of calibrated solar models. 

Equations (18) and (19) of MP give diffusion velocities of trace elements in a H-He background. The deviation between our exact results and the MP and TBL approximations is even worse for heavy element diffusion, as can be seen for oxygen in the right panel of Fig.~\ref{4wtot}.

\section{Progenitors of subdwarf B stars}\label{4rgb}
We distinguish two types of sdB stars: the canonical post-flash sdB star and the newly proposed post-non-degenerate sdB star \citep{han2002, hu2007, hu2008}\footnote{We do not consider sdBs that are merger products.}. The latter type originates from an intermediate mass progenitor ($2 M_{\odot}\lesssim M_{\rm ZAMS} \lesssim 4 M_{\odot}$) that ignited He quiescently, whereas the progenitor of the canonical sdB star is a low mass star ($M_{\rm ZAMS}\lesssim 2$ M$_{\odot}$). To become an sdB star, a red giant must have lost a lot of mass when the core was close to He ignition. For a single star this requires an enhanced stellar wind \citep{dcruz1996}. Since the majority of sdBs are observed in binaries (e.g.~\citealt{maxted2001,morales2006}), a more natural mechanism for the mass loss is binary interaction, which can be either stable Roche lobe overflow or common-envelope ejection \citep{han2002}. Regardless of the mechanism, we just removed the desired amount of envelope to construct an sdB {model}. 

As representative sdB progenitors, we use a 1 M$_{\odot}$ and a 3 M$_{\odot}$ stellar model that we evolved from the ZAMS to the tip of the RGB. At 'zero-age' the 1 M$_{\odot}$ model is assumed to be chemically homogeneous, while in the 3 M$_{\odot}$ model  $^{12}$C has reached equilibrium through the CNO cycle. We used quasi-solar abundances, $X=0.70$, $Y=0.28$, $Z=0.02$. 

For comparison we computed for each stellar mass two evolutionary series, one  with atomic diffusion as described further in Section \ref{4diffusion}, and one without. In the first case, all elements are diffused from the ZAMS to the RGB tip.

\subsection{Low mass progenitor }
One of the main differences between the 1 M$_{\odot}$ {model} with and without atomic diffusion is the difference in the He-core mass at the He-flash, a fact that was already noticed by \citet{michaud2007}. We find that, with atomic diffusion, the red giant has a $M_{\rm core, tip} =0.4664$ M$_{\odot}$, while without atomic diffusion we get $M_{\rm core, tip} = 0.4649$ M$_{\odot}$, where we defined the He core boundary to be at $X=0.10$. This $0.0015$ M$_{\odot}$ difference is caused by the fact that, due to gravitational settling, the He-abundance at the H-burning shell will be lower. This implies a smaller mean molecular weight at the shell for the diffusive model. Thus, the nuclear burning rate will be lower, and hence the maximum temperature in the core increases more slowly.  Consequently, the core mass can grow to a higher value before the temperature for He-ignition is reached. We find $\Delta M_{\rm core. tip}/\Delta Y_{\rm shell} = -0.17$, a somewhat lower dependence than found by  \citet{rood1972} and \citet{sweigart1978}, -0.23 and -0.24, respectively. This could be due to the fact that in our models the difference in He-abundance grows gradually, while  \citet{rood1972} and \citet{sweigart1978} examined the effect of the initial composition using evolution calculations without diffusion. Also, different input physics such as the nuclear reaction rates might play a role.

In a previous study, \citet{hu2007} found a He core mass of $0.472$ M$_{\odot}$ at the tip of the RGB for a 1 M$_{\odot}$ {model}. The current lower value is caused by the new conductive opacities by \citet{cassisi2007}. These authors already mention that their updated conductive opacities cause a different thermal stratification in the He core than previous results, leading to a lower He mass at the onset of the flash. 

\begin{figure}
   \begin{center}
  \includegraphics[angle=-90, width=9.5cm]{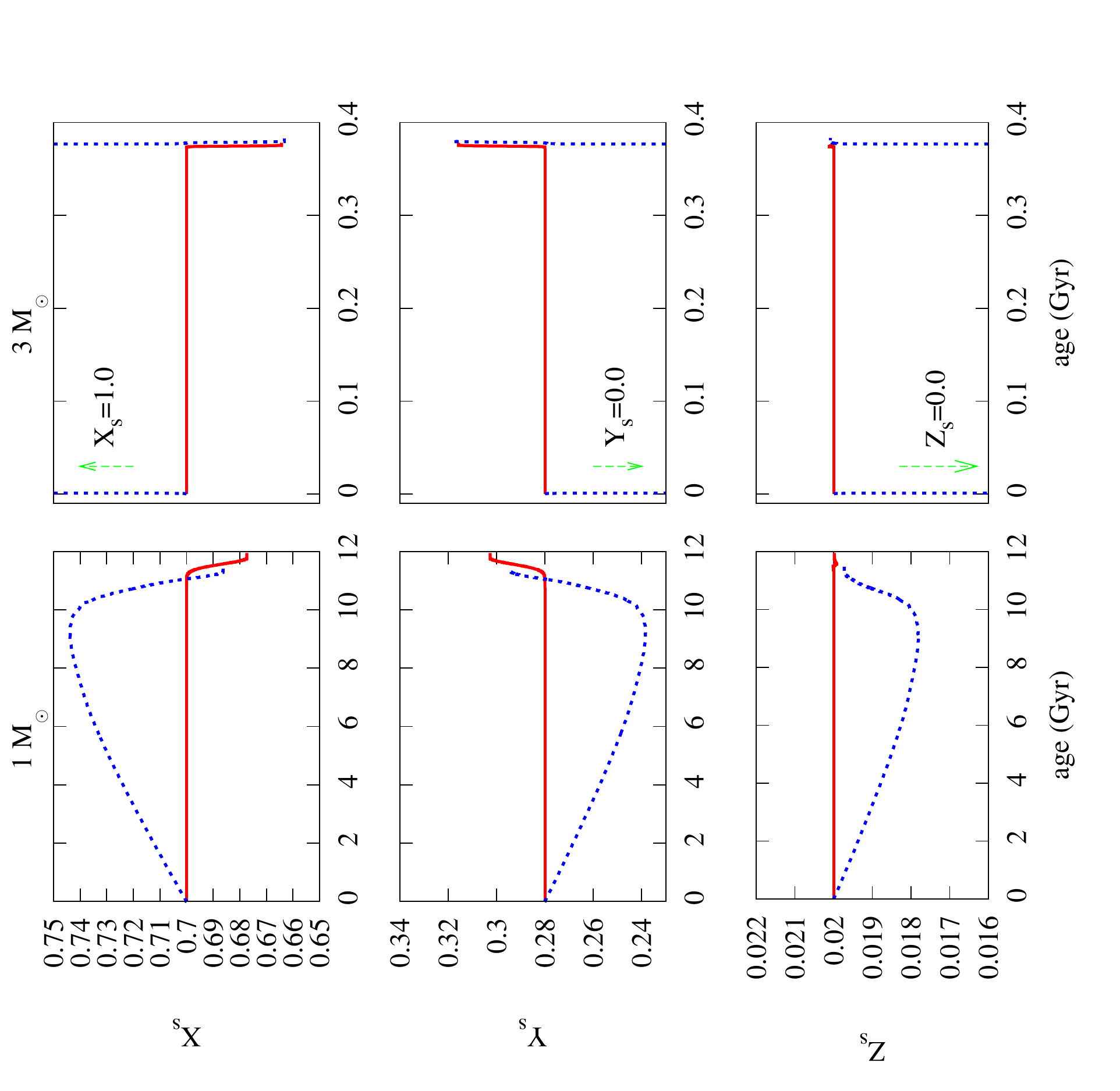}
   \caption{Surface abundances from the ZAMS to the tip of the RGB as a function of the stellar age. The dashed lines are for the model with atomic diffusion, and the solid lines are for the model without. The \emph{left} panels are for the {model}  with $M_{\rm ZAMS}=1$~M$_{\odot}$ and the \emph{right} panels for $M_{\rm ZAMS}=3$~M$_{\odot}$. {For the diffusive  $M_{\rm ZAMS}=3$ M$_{\odot}$ {model} , the abundances run off the scale to the values indicated with arrows.} }
              \label{4surface}
              \end{center}
    \end{figure}

We also notice small, but significant, differences in the surface composition, see Fig.~\ref{4surface}. Even after the first dredge up some differences remain. At the tip of the RGB, the He surface abundance of the diffusive model is 3\% lower than that of the non-diffusive model. The metallicity is lower by 1\%. 

\subsection{Intermediate mass progenitor}
For the 3 M$_{\odot}$ {model} with diffusion, the surface abundances become unrealistic. The outer layers consist of pure H almost immediately after the ZAMS, because the envelope is radiative. In reality, this would be prevented by competing processes such as radiative levitation, turbulence, rotational mixing, mass loss etc. However, we are not worried by this as we will remove most of the envelope to construct an sdB star. In any case, after the first dredge up, the surface abundances become comparable to the model without diffusion. This is partly because diffusion has not been efficient in the interior, during the much shorter evolutionary timescale of the 3 M$_{\odot}$ {model} . The 1 M$_{\odot}$ {model}  evolved from the ZAMS to the RGB tip in $1.2\times 10^{10}$ yr, while it took the 3 M$_{\odot}$ {model}  only $3.8\times10^{8}$ yr. Furthermore, the convective envelope of the 3 M$_{\odot}$ {model}  dredges up more nuclear processed material than in the case of the 1 M$_{\odot}$ {model} , washing away the effects of diffusion at the surface. 

We find, with diffusion, $M_{\rm core,tip} = 0.4403$ M$_{\odot}$, and without $M_{\rm core,tip} = 0.4367$ M$_{\odot}$. The difference is 0.0036 M$_{\odot}$, which is larger than for the 1 M$_{\odot}$ model. At first instance this might be surprising, but one should realize that the reason that the diffusive model has a larger $M_{\rm core,tip}$ is different than for the 1 M$_{\odot}$ model. In this case, the convective core  on the MS becomes larger in the presence of diffusion. This is mainly caused by chemical composition changes leading to an increase in the opacity as shown by \citet{richard2001} and \citet{michaud2004}. An extended convective core can burn more H, which leads to a larger He core at the end of the MS, and ultimately, to a larger He core at the RGB tip.

\section{Subdwarf B models}\label{4sdb}
During the evolution up to the RGB tip, all elements were diffused if diffusion was taken into account. However, on the EHB we must take special care because in the absence of competing forces gravitational settling will cause all heavy elements to sink. The outer layers will consist of pure H, and we know this is not true for sdB stars. Their atmospheres are usually He deficient, a typical value is $\log [N(\mbox{He})/N(\mbox{H})]=-2$, although it can vary from $-4$ to $-1$ \citep{saffer1994, heber2004}. The metal  abundances show a wide spread from solar to subsolar by a factor 100, and different metals show different patterns, see e.g.~\citet{fontaine2004} and \citet{otoole2006}. Current diffusion theory, with radiative levitation and weak stellar winds, has difficulties explaining the observed abundance anomalies. We make no such attempt here, but we evaluate 
how our results are affected by certain assumptions about the surface abundances. We examine five cases on the EHB in order to disentangle the impact of different diffusion processes:
\begin{itemize}
\item[1a)] No diffusion, Fe is not enhanced.
\item[1b)] No diffusion, Fe is enhanced.
\item[2a)] Only H and He are diffused, so the surface metallicity stays roughly solar ($Z\approx 0.02$). Fe is not enhanced.
\item[2b)] Only H and He are diffused, Fe is enhanced. 
\item[2c)] All elements are diffused, 
Fe is not enhanced.  
\end{itemize}
We note that the differences between case 1a) and 1 b) were already examined in \citet{hu2008}.
In principle, we expect that case 2b) is the most realistic, because diffusion calculations in HB stars indicate that He settles while the metals are supported by radiative levitation (\citealt{michaud1983, michaud2008}). It is a tentative conclusion though, since it is not clear to what extend the results for HB stars apply to sdB stars that have higher  surface gravities and effective temperatures. Also, these authors examined HB stars with Z=0.0001, and radiative accelerations in a solar metallicity star are much smaller due to saturation of the lines. Case 2c) tells us what happens if radiative levitation is not effective.

\subsection{Post-flash sdB stars}

   \begin{figure*}
   \begin{center}
  \includegraphics[angle=-90, width=18cm]{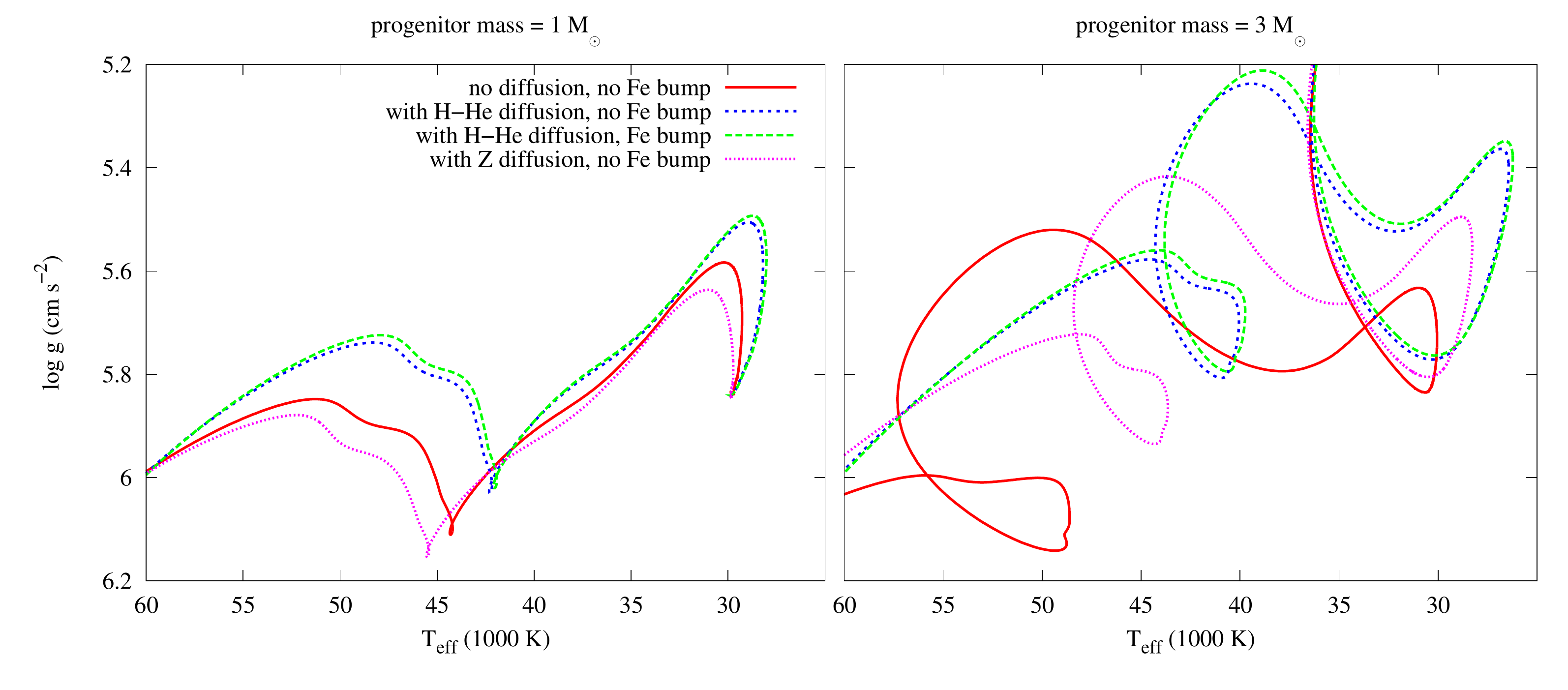}
   \caption{Evolutionary tracks for \emph{left}: the post-flash sdB star, and \emph{right}:  the post-non-degenerate sdB star. The lines correspond to the cases 1a) solid line, 2a), short-dashed line, 2b) long-dashed line and 2c) dotted line as described in the text.}
              \label{4evtracks}
              \end{center}
    \end{figure*}
   \begin{figure*}
   \begin{center}
  \includegraphics[angle=-90, width=18cm]{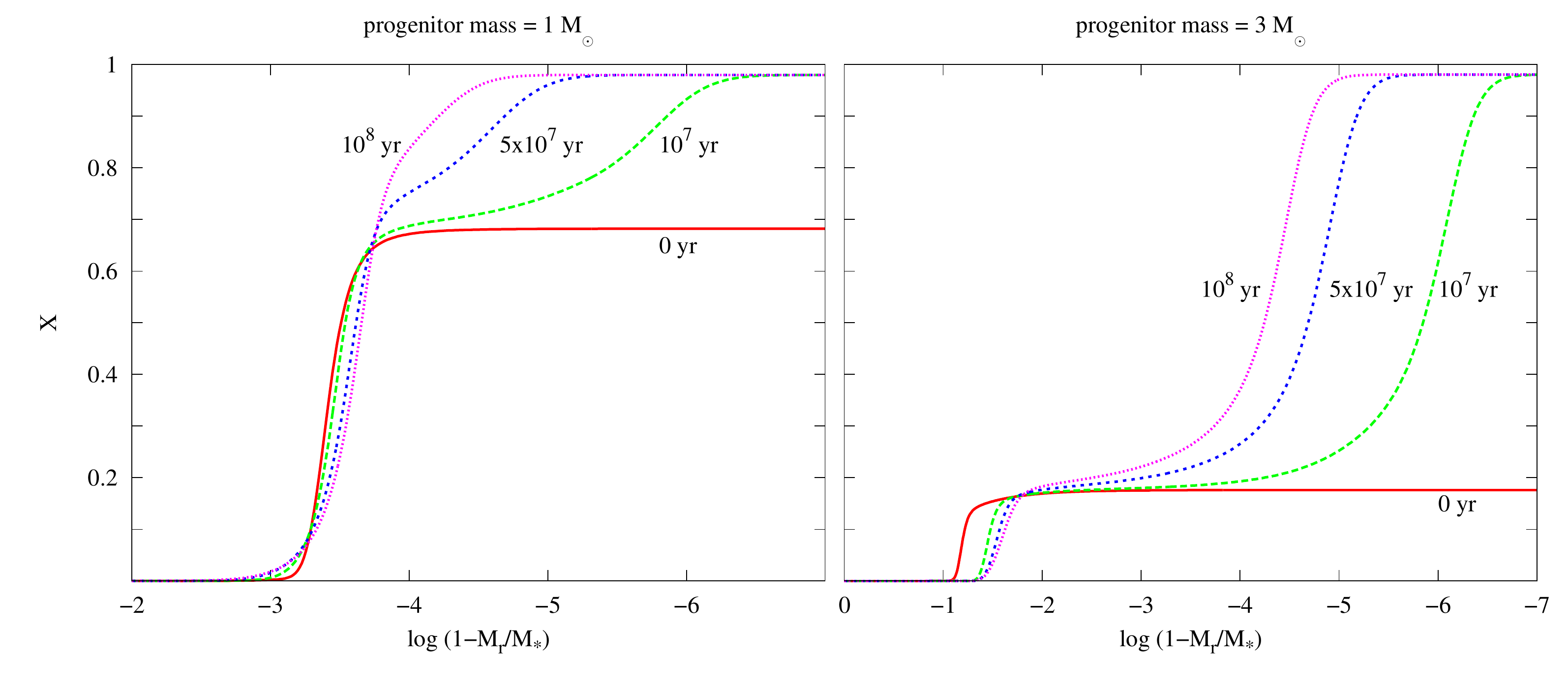}
   \caption{{The effect of diffusion on the the H-profiles at different EHB ages for \emph{left}: the post-flash sdB star, and \emph{right}: the post-non-degenerate sdB star.}}
              \label{4diffprofile}
              \end{center}
    \end{figure*}

The post-flash sdB models are constructed from the 1 M$_{\odot}$ series. If the luminosity due to He-burning exceeds $10^4$ L$_{\odot}$, it is assumed that the He-flash occurs. The post-flash model is taken to be a ZAHB star with the same mass, core mass and chemical composition as the pre-flash model. In reality, the core abundances  are changed slightly due to nuclear burning during the flash.  Detailed calculations of the flash by \citet{piersanti2004} and \citet{serenelli2005}, suggest that C increases up to 5\%. We are not concerned about this, since the p-modes are shallow envelope modes and do not probe the core. However, for the g-modes that propagate to deeper interior regions, the effects of nuclear burning and convective mixing during the flash (as described in e.g. \citealt{dearborn2006} and \citealt{mocak2008}) {are} important. {We explored  this in \citet{hu2009}}.

For the series without diffusion, the H-envelope is removed just after He-ignition. The compositions are kept fixed during mass loss, i.e. it is assumed that the mass loss happens on a timescale shorter than the nuclear and diffusion timescales. The total amount of H left is  $M_{\rm H}=10^{-4}$ M$_{\odot}$. In this way, we obtained sdB models with $M_{\rm core} = 0.465$ M$_{\odot}$.  To obtain sdB models with the same core mass for the series with diffusion, we had to remove the envelope before He-ignition. {Still, the core mass is large enough so that He is ignited after mass-loss, see \citet{hu2007}.}

In the left panel of Fig.~\ref{4evtracks}, we show the sdB evolutionary tracks in the $\log g-T_{\rm eff}$ diagram for the cases 1a) solid line, 2a) short-dashed line, 2b) long-dashed line and 2c) dotted line. 
We observe the following:
\begin{itemize}
\item  Compared to 1a) \emph{no diffusion, no Fe enhancement}, track 2a) \emph{with H-He diffusion and no Fe enhancement}, is shifted to lower surface gravities and effective temperatures. Due to the outward diffusion of H, the density in the envelope decreases. Consequently the envelope becomes less gravitationally bound, and the sdB star gets larger and cooler. 
\item Comparing track 2a) to 2b) \emph{with H-He diffusion and with Fe enhancement} shows that Fe enhancement has only a small effect on the sdB evolution as we already 
saw in \citet{hu2008}. Due to the increased opacity caused by Fe enhancement there is a slight shift to lower gravities and temperatures.
\item Interestingly, track 2c) \emph{with all elements diffusing, and no Fe enhancement}, is shifted to higher gravities and temperatures compared to 1a). This is because the opacity decreases as the heavy elements sink. Consequently the star becomes more compact and hotter, and apparently this effect is greater than what happens due to H-He diffusion in 2a).
\end{itemize}

To illustrate the effect of time-dependent diffusion, we show in Fig.~\ref{4diffprofile}, the H-profile against the fractional mass $\log(1-M_r/M_*)$ at four different ages of the sdB star: $0$, $10^7$, $5\times10^7$, and $10^8$ yr after the ZAEHB.  Without diffusion the H-profile would hardly change.

\begin{figure}
   \begin{center}
  \includegraphics[angle=-90, width=9cm]{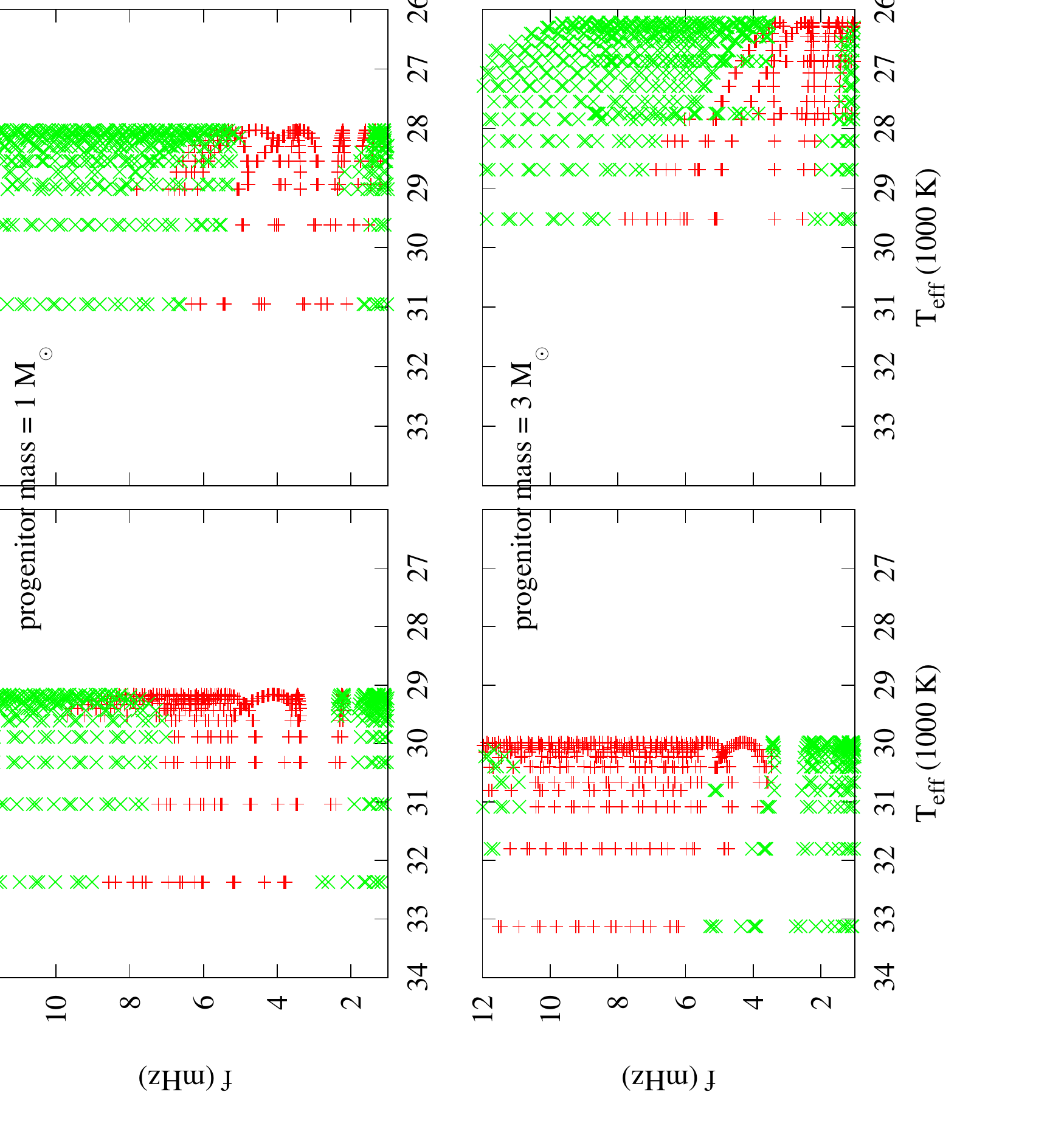}
   \caption{{The frequencies of the stable ($\times$) and unstable ($+$) modes with $l\leq2$ as a function of the effective temperature. The \emph{upper} panels are for the post-flash sdB star and the \emph{lower} ones are for the post-non-degenerate sdB star. We used the structure models from evolutionary tracks 1b) and 2b) for the case of no diffusion (\emph{left}) and with diffusion (\emph{right}), respectively.}}
              \label{4mode}
              \end{center}
    \end{figure}

After each $10^7$ years of evolution on the EHB, the pulsational properties were calculated with {the non-adiabatic stellar oscillation code MAD \citep{dupret2001}}. In the top panels of Fig.~\ref{4mode}, we show the frequencies of pulsation modes  $0\leq l \leq 2$ as a function of the {effective temperature}. We plotted the results for the tracks with Fe enhancement in order to evaluate the mode excitation, namely 1b), \emph{no diffusion, with Fe enhancement}, and 2b), \emph{H-He diffusion, with Fe enhancement}. We notice  distinct changes in both the frequencies and the frequency range of excited modes.

To understand the latter, one must realize that the driving is caused by an opacity bump at $\log T =5.3$. When H diffuses outwards, the sdB star is cooler and has a lower temperature gradient due to the lower density in the envelope by
\begin{equation}\label{radtransport}
\frac{dT}{dr}=-\frac{3}{4ac}\frac{\kappa\rho F}{T^3}.
\end{equation}
It should be noted that an increase in the H-fraction tends to increase the opacity, which also affects the radiative energy transport  Eq.~(\ref{radtransport}). However, the work of \citet{jeffery2006a} indicates that the effect of reducing the mean molecular weight outweighs the effect of  increasing the opacity, and our results here regarding the mode excitation support their findings. Thus the opacity bump will be located deeper in the star in the presence of H-He diffusion.
For a mode to be driven,  the amplitudes of the eigenfunctions must be significant in the driving region, which occurs when the last node is at a certain temperature ($\log T\approx6$, see \citealt{hu2008} for details). So the last node of a mode must also be located deeper in the star in order to get excited. This corresponds to modes of lower radial order and frequency.

\subsection{Post-non-degenerate sdB stars}
For this type of sdB stars, we used the $3$ M$_{\odot}$ model as progenitor. The envelope is removed just after He ignited quiescently in the core until the total mass is 0.465 M$_{\odot}$. 

In the right panel of Fig.~\ref{4evtracks}, we show the sdB evolutionary tracks in the $\log g-T_{\rm eff}$ diagram for the same cases as before. Here, we deduce that;
\begin{itemize}
\item Compared to 1a), \emph{no diffusion, no Fe enhancement}, track 2a), \emph{with H-He diffusion and no Fe enhancement}, is drastically shifted to lower gravities and effective temperatures. In the presence of diffusion the initial low H-abundance ($X_s=0.18$) in the sdB envelope cannot be maintained. The post-non-degenerate sdB star has a lot of H 'hidden' in deeper layers. When this diffuses outwards, the resulting shift in the $\log g-T_{\rm eff}$ diagram is much more pronounced than for the post-flash sdB star. 
\item Again we find that Fe enhancement shifts track 2b), \emph{with H-He diffusion and with Fe enhancement}, to slightly lower gravities and effective temperatures compared to 2a). 
\item Track 2c), \emph{with all elements diffusing, and no Fe enhancement}, shows an interesting feature. It is shifted to higher gravities and effective temperatures compared to track 2a), but still has lower gravities and temperatures compared to 1a). This is in contrast to what happens in this case for the post-flash sdB star. It is apparent that the decrease in opacity has a smaller impact than the effect of  H-He diffusion described in 2a). 
\end{itemize}

In the right panel of Fig.~\ref{4diffprofile}, one can see the effect of H-He diffusion on the H-profile.  The abundances in the outer layers are changed on a short timescale. At  depths of $\log(1-M_r/M_*)>-3$, diffusion does not work efficiently on the EHB timescale ($\sim 10^8$ yr). The change in the H-abundance near the core at $\log(1-M_r/M_*)\gtrsim-1$, is caused by H-shell burning.

In the bottom panels of Fig.~\ref{4mode}, we plot the frequencies of pulsation modes  $0\leq l \leq 2$ {against the effective temperature for the cases 1b) and 2b)}. The effects of diffusion on the pulsations are {distinct}. This is not surprising considering the great impact diffusion has on the structure of the post-non-degenerate sdB star. A previous study \citep{hu2008} found that the frequency range of excited modes is one of the main discriminators between the post-flash and the post-non-degenerate sdB star. {If diffusion is effective}, this conclusion must be revised. We still see differences in the frequency spectrum between models from different evolutionary channels, but a more detailed study including more models is necessary to quantify the differences. {We plan to perform such an analysis in the near future.}

We conclude that  atomic diffusion causes a larger change of H surface abundance in the post-non-degenerate sdB star than in the post-flash one. So the impact of H-He diffusion on the evolutionary tracks and driving is larger. In contrast to the post-flash case, this cannot be compensated by a possible decrease in the opacity by the heavy elements settling.

\section{Conclusions}\label{4concl}
We have updated the diffusion routine by TBL to make use of diffusion coefficients derived from a screened Coulomb potential \citep{paquette1986} rather than a pure Coulomb potential. The improved accuracy is mainly in the thermal diffusion term, and this results in slightly lower diffusion velocities compared to TBL. Although the effect is small, it becomes important within the accuracy desired for calibrated solar models. We found an excellent fit to the Sun by evolving a 1 M$_{\odot}$ with initial abundances $Z=0.01928$, $Y=0.2725$ and mixing length parameter $\alpha=2.1$ with atomic diffusion to the age of $4.57\times 10^9$ yr. 

We evolved a 1 and 3 M$_{\odot}$ {stellar model}  from the ZAMS to the RGB tip, with and without atomic diffusion. By including diffusion, we find an increase in the He core mass at the RGB tip of 0.0015 M$_{\odot}$ and 0.0036 M$_{\odot}$, respectively.  Surface mass fractions are slightly changed for the 1 M$_{\odot}$ model, but not significantly for the 3 M$_{\odot}$ model. The reason is in part the shorter evolutionary timescale of the higher mass star. More importantly though, is the more efficient mixing of nuclear processed material during the first dredge up. This is because for the higher mass star, the inward advance of the convective envelope reaches regions where the shrinking convective core  passed through during the main sequence.

From the RGB models, we constructed post-flash and post-non-degenerate sdB models. Radiative levitation was not included, but in one set of our calculations we enhanced Fe artificially around $\log T=5.3$ as an approximation to the expected diffusive equilibrium profile \citep{charpinet1996}. This allows us to study the excitation of the pulsation modes, while the effect of such an artificial Fe profile on the sdB evolution is minimal. Although the question remains if such models are suitable for a precise seismic analysis, they are very useful for a comparative study as presented here.

First of all, we see significant shifts in the evolutionary tracks when H-He diffusion is included. If H diffuses outwards, the envelope is less dense and thus less gravitationally bound. This results in larger radii, and therefore lower gravities and effective temperatures. For the post-non-degenerate sdB star the difference is much more pronounced, because the initial (ZAEHB) H-abundance in the envelope was very low, $X=0.18$, and H extended to deeper layers. In the presence of H-He diffusion in our sdB models, the frequencies of excited modes are lower and the frequency spectrum is more densely spaced. {Especially for the post-non-degenerate  sdB star the effect is drastic due to the large structural change of the envelope. } 

We also examined what happens if the metals are allowed to sink due to gravitational settling. The corresponding decrease in opacity will tend to make the star {more compact} and hotter, thus competing with the effect of He settling. In reality, however, radiative levitation prevents the metals from sinking, so we expect the case of only H-He diffusion to be more realistic. It is clear that consistent modelling of radiative levitation is a missing piece in this study, and we intend to include this in future work. Still, the results presented here, are an improvement to previous sdB models that altogether neglect atomic diffusion. In particular, we find it cannot be ignored in the post-non-degenerate sdB stars as it leads to totally different stellar structures. {Furthermore, for H and He, the radiative forces do not play a significant role compared to settling. Therefore, the main conclusion of our work remains valid even when other processes are included: The settling of He leads to a H-rich sdB envelope, which leads to a lower moleculair weight.  This results to a lower density at the same pressure and temperature, which changes both the evolution tracks and the stability of the modes in the way presented here.}

\begin{acknowledgements}
The authors are indebted to R.~Scuflaire and O.~Pols for their help on the opacity tables. HH thanks S.~Kawaler and L. Yungelson for helpful discussions and suggestions.  
HH acknowledges a PhD scholarship through the ``Convenant
Katholieke Universiteit Leuven, Belgium -- Radboud Universiteit Nijmegen, the
Netherlands''. GN is supported by NWO VIDI grant 639.042.813. 
The research leading to these results has received funding from the
European Research Council under the European Community's Seventh Framework Programme (FP7/2007--2013)/ERC grant agreement n$^\circ$227224 (PROSPERITY), as well as from the Research Council of K.U.Leuven grant agreement GOA/2008/04. RJS is funded by the Australian Research Council's Discovery Projects
scheme under grant DP0879472.

\end{acknowledgements}

\bibliographystyle{aa}
\bibliography{12290}
\end{document}